\pdfoutput=1
\documentclass[reprint,amsmath,amssymb,aps,nofootinbib]{revtex4-1}
\usepackage{graphicx}% Include figure files
\usepackage{dcolumn}% Align table columns on decimal point
\usepackage{bm}% bold math
\usepackage{mathrsfs}
\usepackage{ascmac}
\usepackage{latexsym}
\usepackage{amsmath}
\usepackage{here}
\usepackage{setspace}
\usepackage{color}
\usepackage{setspace}
\def\ve{\varepsilon}
\def\ver{\varepsilon_R}
\def\hlf{\frac{1}{2}}
\def\vsp{\noalign{\vskip 0.3cm}}
\def\vssp{\vspace{0.2cm}}
\def\bea{\begin{eqnarray}}
\def\eea{\end{eqnarray}}
\def\begenu{\begin{enumerate}}
\def\endenu{\end{enumerate}}
\def\bit{\begin{itemize}}
\def\eit{\end{itemize}}
\def\hlf{{1\over 2}}
\def\ndim{\frac{d^n \ell}{i(2\pi)^n}}
\def\psix{\left(\frac{1}{16\pi^2}\right)}
\def\delW{\delta Z_W^{1/2}}
\def\delA{\delta Z_{AA}^{1/2}}

\def\delZA{\delta Z_{ZA}^{1/2}}
\def\delAZ{\delta Z_{AZ}^{1/2}}
\def\rar{\rightarrow}
\def\xnla{\tilde{\alpha}}
\def\xnlb{\tilde{\beta}}
\def\xnlk{\tilde{\kappa}}
\def\xnld{\tilde{\delta}}
\def\xnle{\tilde{\varepsilon}}
\def\m{M^2}
\def\ve{\varepsilon}
\def\mz{M_Z^2}
\def\mw{M_W^2}
\def\gmn{g_{\mu\nu}}
\def\qmn{q_\mu q_\nu}

\def\mhs{\mathchar`-}
\def\nn{\nonumber}
\begin{document}
\title{Numerical calculation of the full two-loop electroweak corrections  to muon ($g$-2)
}

\author{ Tadashi Ishikawa}
\email{tishika@suchix.kek.jp}
\affiliation{High Energy Accelerator Organization(KEK), 1-1 OHO Tsukuba Ibaraki 305-0801,Japan}
\author{Nobuya Nakazawa}%
 \email{nobuya@suchix.kek.jp}
\affiliation{Department of Physics, Kogakuin University, Shinjuku,Tokyo 163-8677,Japan}
\author{Yoshiaki Yasui}
\email{yasui@tokyo-keitan.ac.jp}
\affiliation{Department of Management, Tokyo Management College,Ichikawa, Chiba 272-0001, Japan}

\date{\today}% It is always \today, today,

\begin{abstract}
Numerical calculation of two-loop electroweak corrections to 
the muon anomalous magnetic moment ($g$-2)
is done based on, on shell renormalization scheme (OS) and free quark model (FQM).
The GRACE-FORM system is used to generate Feynman diagrams and corresponding amplitudes.
 Total 1780   two-loop diagrams and 70 one-loop diagrams composed of  counter terms are calculated
to get the renormalized quantity. 
As for the numerical calculation, 
we adopt trapezoidal rule with Double Exponential method (DE). 
Linear extrapolation method (LE) is introduced 
to regularize UV- and IR-divergences and to get finite values.  
The reliability of our result is guaranteed by several conditions.
The sum of  one and  two  loop electroweak corrections in this renormalization scheme
becomes $a_\mu^{EW:OS}[1{\rm+}2\mhs {\rm loop}]= 151.2 (\pm 1.0)\times 10^{-11}$, where the
error is due to the numerical integration and the uncertainty of input mass parameters 
and of the hadronic corrections to electroweak loops.
By taking the hadronic corrections into account, we get
 $a_\mu^{EW}[1{\rm+}2\mhs {\rm loop}]= 152.9 (\pm 1.0)\times 10^{-11}$.
It is in agreement  with the previous works given in PDG\cite{PDG} within
errors.
 
\end{abstract}
%%%%%%%%%%%%%%%%%%%%%%%%%%%%%%%
\pacs{Valid PACS appear here}% PACS, the Physics and Astronomy
\maketitle
%%%%%%%%%%%%%%%%%%%%%%%%%%%%%%
\section{\label{sec:Intro}Introduction}
%%%%%%%%%%%%%%%%%%%%%%%%%%%%%%
In order to get a sign of beyond the standard model
physics from high precision experimental data, 
 we need  higher order radiative corrections within Standard Model~(SM). For this purpose our group
 has been developing the  automatic calculation system GRACE [\onlinecite{GRACE}] since the late 1980's.
 The measurement of the muon anomalous magnetic moment $a_\mu \equiv (g\mhs 2)/2$  is 
 the one of the most precise experiments to 
 check the SM. QED correction was calculated by T.Kinoshita et al.[\onlinecite{Kinoshita}] up to tenth-order.
 The two-loop electroweak (ELWK) correction to $a_\mu$  was calculated approximately by  
 Kukhto et al.[\onlinecite{Kukhto}] in 1992.
 Surprisingly,  the two-loop correction is almost 20\% of the one-loop correction 
 [\onlinecite{Weak1,Weak1b,Weak1c,Weak1d}]. 
 We started to calculate the full two-loop corrections in 1995 and presented
 our formalism at Pisa conference [\onlinecite{Pisa}].  We also showed that  
 the two-loop QED value 
[\onlinecite{Karplus,QED-Som1,QED-Peter,QED-Som2}]
 was correctly reproduced within our general formalism. 
 However, the number of diagrams is huge 
  and the numerical  integration 
 requires the big CPU-power to achieve required accuracy, 
 we must wait until various environments are improved. 
 
 During these days, the several groups did the 
 approximate calculations
  [\onlinecite{Marciano,Marciano2,Marciano3,Grib,Weak2,Heine}]
 and the approximate value of the two-loop ELWK correction is widely accepted [\onlinecite{PDG},\onlinecite{Marquard}].
 In 2001, BNL-Experiment 821
  [\onlinecite{BNL1},\onlinecite{BNL2}] announced that the precise experimental 
 value deviates from that of SM around $(2.2 \sim 2.7) \sigma~$[\onlinecite{BNL3}]. 
 It brought  much interest in the theoretical
 value. 
 The main theoretical concern is now shifted to the hadronic contributions [\onlinecite{Davier},\onlinecite{Prades}].
However, the discrepancy between the experimental value and 
the theoretical value is still large $\sim 3.5 \sigma$ [\onlinecite{PDG},\onlinecite{CODATA}]. 
As new experiments
at FNAL-E989 [\onlinecite{E989}] will announce their first result in 2019 
and J-PARC-E034 [\onlinecite{Jparc}] is also planning the new experiment, 
we can expect to have a new data soon.
%%%%%%%%%%%%%%%%%%%%%%
\subsection{\label{sec:PNQFT} Perturbative Numerical QFT}
Although the two-loop ELWK correction is almost established,
we try to get the value without any approximation to confirm 
the validity of the earlier studies
\footnote{An intermediate stage of our calculation was reported in [\onlinecite{CPP2016}]}.
This work is an important milestone  to extend GRACE-system 
from one-loop to  two-loop calculation.  
In ELWK theory,  there are so many fields, mass parameters and 
complex couplings that it is hard to get reliable higher order corrections  
to  physical quantities, in general. 
%%%%%%%%%%%%%%%%%%%%%%%%%%%%%%%%%%%%
It is desirable to construct the framework to calculate these
higher order corrections as automatically as possible.
The key point is to perform Feynman integration numerically 
by using a sophisticated method with good convergence and 
high power CPU machine. 
We propose to call  such  framework as   
Perturbative Numerical Quantum Field Theory (PNQFT).
The concepts of PNQFT  are  based on the following principles.
\begin{description}
\item[(a)] It is essential to assume  amplitudes as meromorphic functions
of space time dimension $n$ for regularization and getting 
 gauge invariant  renormalized values of physical quantities.
\item[(b)] The source  program  for numerical integration is  automatically generated 
by GRACE together with a symbolic manipulation system such as FORM [\onlinecite{FORM}].
\item[(c)]  A high precision numerical integration method should be adopted.
\item[(d)] Linear Expansion method (LE) (see subsection \ref{subsec:Reg}) [\onlinecite{LEM},\onlinecite{Donc224}]
is crucial to  extract 
both UV- and IR-divergences by taking advantage of the above analyticity. 
By LE method, we can expand the amplitude 
in any order of $\ve(=2-n/2)$ (Laurent expansion), so that it is a powerful 
tool  for higher order calculation.  
\item[(e)] To guarantee the validity of the calculation, several conditions must
be cleared. An example is the cancellation of non-linear gauge (NLG) parameters.
\item[(f)]  It is crucial to reduce human intervention to avoid careless mistakes.
We must  minimize the handmade operations necessary for getting the physical quantities.    
\end{description}   
The following calculation is based on these principles.
In section \ref{sec:Outline} and \ref{sec:Logic}, we briefly explain the flow and framework of our 
calculation.
In section \ref{sec:Numcal}, we touch on our method of numerical calculation.
We emphasize that the Linear Extrapolation (LE) method is simple and efficient method
to regularize UV- and IR-divergences and also to get finite values.
We also explain  our consistency conditions to ensure the results.  
Some examples of calculations are explained.
In section \ref{sec:Result}, we give  our results on $a_\mu$ .
In the last section, we give some comments to make extensive progress.
In Appendices, we explain the technical parts of our calculation.
%%%%%%%%%%%%%%%%%%%%%%%%%%%%%%
\section{\label{sec:Outline}Outline of our frame work}
\begin{figure}[h]
\includegraphics[scale=0.27,bb=0 0 934 581]{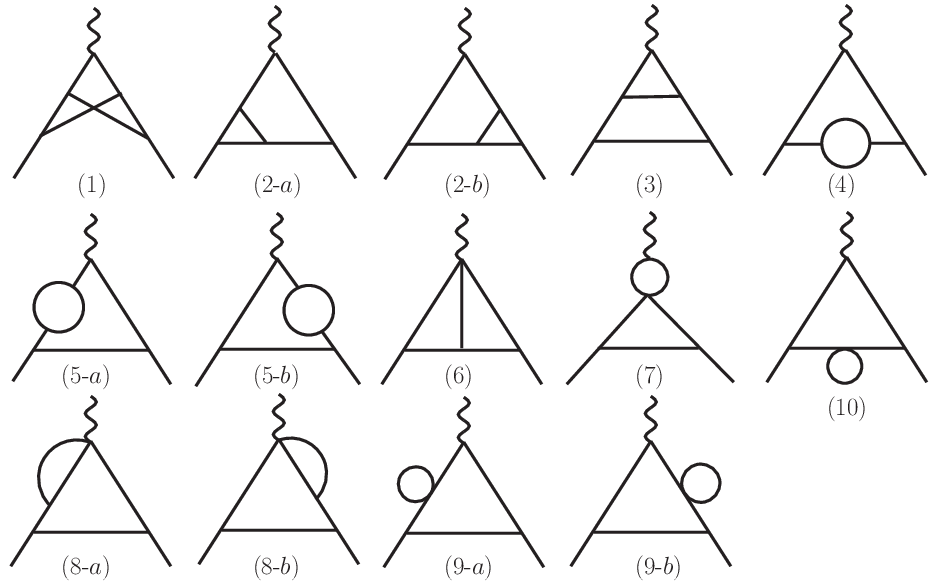}
\caption{\label{fig:topol} Types of topology}
\end{figure}

%%%%%%%%%%%%%%%%%%%%%%%%%%%%%%
Our calculation is formulated under the following conditions.
\begenu
\item
The calculation is done within SM.
\item
On mass shell renormalization scheme  (OS)  is adopted 
[\onlinecite{RenormMuta,RenormPTP100,Grace2}].
We adopt $\alpha, M_Z, M_W, M_H$  and fermion masses as physical 
parameters. Weinberg angle, Higgs fermion coupling  and other quantities 
are expressed by  these parameters. 
\item 
Free quark  model (FQM) is adopted and  as for quarks, constituent masses
are used. 
\item
Non-linear gauge  formulation with 't\,Hooft-Feynman propagator is adopted.
\item
Dimensional regularization is applied for both Ultra Violet (UV)- and Infrared (IR)-divergences.
\item
Linear Extrapolation method (LE) is fully used for regularization and 
getting finite values.
\endenu
Next, we briefly explain the flow of our calculation.
\begenu
\item
GRACE system generates all the diagrams we need in SM, automatically [\onlinecite{KanekoFG}].
There are 1780 two-loop 
diagrams\footnote{There are 1678 diagrams in Feynman gauge and 102 extra diagrams specific to NLG}
and 70  one-loop diagrams
composed of  one-loop order counter term (CT).
Two-loop order CT is not necessary in our case, because 
$a_\mu$  is not related to the charge renormalization part.
\item
These 1780 diagrams are classified into 14 types of topology.
Types of  the topology are displayed 
in Fig.\ref{fig:topol}. 
Among these types, some of them give the same contribution
because of  symmetry. (an example: 5-a vs. 5-b)
 The diagrams including CT
are classified into two types, namely, vertex and self-energy types.\\
%%%%%%% Figure: Fig-1  %%%%%%%%%%%
%%%%%%%%%%%%%%%%%%%%%%%
\item
GRACE system generates the amplitude of each diagram
in accordance with  Feynman rules for 
ELWK theory with NLG [\onlinecite{Grace2}].
 \item
In order to express the amplitude as a function of Feynman parameters
used to combine denominators, we define the following quantities for each topology 
in advance [\onlinecite{Cvita},\onlinecite{Cvita2}].
\bit
\item
Internal loop momentum flow$(\ell_s,\eta(s))$ ($s=1,2$)
\item
External momentum flow $(q_j)$   
($j=$ internal line number)
\item
Kirchhoff's law of momentum conservation at each vertex
\item
Feynman parameters are transformed to the integration variables 
in the interval [0,1].
\eit
\item
Using these tools,  contribution of each diagram 
to $a_\mu$  is expressed as function of Feynman parameters,  
according to the formulas given in the 
next section. 
We make use of a symbolic manipulation system FORM 
exhaustively.
\endenu
%%%%%%%%%%%%%%%%%%%%%%%%
\section{\label{sec:Logic}Logic of the Calculation}
%%%%%%%%%%%%%%%%%%%%%%%%
\subsection{Cvitanovi$\acute{\rm c}$-Kinoshita procedure} 
In order to extract $a_\mu$ factor from muon vertex function,
we adopt Cvitanovi$\acute{\rm c}$-Kinoshita procedure 
[\onlinecite{Cvita},\onlinecite{Cvita2}].
We briefly explain the procedure in the case where there are six propagators.
Starting formula is the two-loop muon vertex,
\bea
\Gamma_\mu 
&=& \int
 \frac{d^n \ell_1}{i(2\pi)^n}
 \frac{d^n \ell_2}{i(2\pi)^n}
 \frac{F_\mu(D)}{\prod_j (p_j^2-m_j^2)}  \nn \\
\vsp
 &=&
 \Gamma(6) \int
 \prod dz_j \delta(1-\sum_j z_j) \nn  \\
 \vsp
 &\times& \int
 \frac{d^n \ell_1}{i(2\pi)^n}
 \frac{d^n \ell_2}{i(2\pi)^n}
 \frac{F_\mu
 (D)}{\sum_j z_j(p_j^2-m_j^2)} 
\eea
 where $\displaystyle{p_j=\sum_{s=1}^2\eta_s(j)\ell_s +q_j}$, is the  momentum on the internal line $(j)$.
The function  $\eta_s(j)(=\pm 1 ,0) $  defines the weight of loop momentum $\ell_s$ 
on the internal line $(j)$. 
The $z_j$'s are the Feynman parameters to combine six propagators.
$F_\mu(D)$ is the numerator function and $\mu$ is the external photon polarization. 
Next  we diagonalize the denominator function with respect to loop 
momenta $\ell_1$, $\ell_2$ and perform 
integration. The result is,
\bea
\Gamma_\mu 
&=&\frac{1}{(4\pi)^n}\int
 \prod dz_j \delta(1-\sum_j z_j)
\frac{\Gamma(6-n)}{({\rm detU})^{n/2}} \nn \\
&\times&F_\mu(D)\frac{1}{(V-i\epsilon)^{6-n}},~~
U_{s,t}=\sum_{j=1}^6z_j\eta_s(j)\eta_t(j)  \label{IR0}. \nn \\
\eea
Where ${U}$ is well known 2$\times$2 matrix, composed of 
Feynman parameters $(z_j)$. 
$V(z_j,m_j, q_j)$ is the denominator function. 
The argument is easily 
extended to the case with five-propagators (diagrams with four-point coupling).  

In order to generate the numerator function we use  the following 
differential integral operator $D^\mu_j$.
\bea
\frac{p_j^\mu}{(p_j^2 - m_j^2)}
=D_j^\mu \frac{1}{(p_j^2 - m_j^2)} ,~
D_j^\mu \equiv  \frac{1}{2}\int\nolimits_{m_j^2}^{\infty}
dm_j^2 \frac{\partial}{\partial q_{j\mu}}   \nn \\
\eea
The operator $D_j^\mu$ generates momentum $p_j^\mu$ on the internal line $(j)$.
By operating $D_j^\mu$ to the denominator function $V$, 
we get the following expression.
\bea
D_i^\mu D_j^\nu \frac{1}{V^m}
&=&\frac{Q_i^\mu Q_j^\nu}{V^m}
+\left(- \frac{1}{2\rm detU}\right)\frac{g^{\mu\nu}}{(m-1)}
       \frac{B_{ij}}{V^{m-1}} \label{2ndD}  \label{TKformula} \\
\vsp
{Q}_j^\mu&=& q_j^\mu-\frac{1}{{\rm det}U } \sum_{i=1}^{6} z_iB_{ij}q_i^\mu \\
\vsp
B_{ij}&=&\sum_{s,t}\eta_s(i)\eta_t(j)U^{-1}_{st}{\rm det }U =B_{ji} 
\eea
Using the above formulas, we can write down the numerator functions
in terms of $B_{ij}$ algebraically.
 The equivalence of the above method and
the well known method of shifting loop momentum 
to diagonalize the denominator function 
is verified. 
The correspondence between two methods are symbolized as follows.
\bea 
\ell^0~&\rar&~ \frac{\{1,Q_i^\mu,Q_i^\mu Q_j^\nu,\cdots\}}{V^m},\nn \\
\ell_i^\mu\ell_j^\nu~&\rar&~~
\left( -\frac{1}{2{\rm detU}}\right)\frac{g^{\mu\nu}}{(m-1)}\frac{B_{ij}}{V^{m-1}}
\eea
Next step is to extract the $a_\mu$  factor by using projection operator,
from the photon muon vertex $\Gamma_\mu$. 
The quantity $a_\mu$  is given as follows. ( $m_0$=muon mass )
\bea
a_\mu &=& \lim_{q^2\rightarrow 0} \frac{{m_0}}{p^4q^2}
 {\rm Tr}\left(\Gamma_\mu {\rm Proj }(\mu) \right)  \nn \\
 \vsp
{\rm Proj}(\mu) &=& \frac{1}{4}(\rlap{/}p-\hlf\rlap{/}q+m_0)
    \{{\rm m_0}\gamma_\mu(p.p)-({m_0}^2+\frac{q.q}{2})p_\mu \} \nn \\
&\times&
    (\rlap{/}p+\hlf\rlap{/}q + {\rm m_0}),~~~ \label{eq:projection}
\eea
where we set momentum of 
incoming $\mu^-$, outgoing $\mu^-$ 
 and incoming photon, as 
$(p-q/2),(p+q/2)$ and $q$, respectively.  \\
The final expression for numerical integration is summarized in the following 
formula.
\bea
F&=& \frac{1}{(4\pi)^n}\int
\prod dz_j \delta(1-\sum_{j} z_j) \nn\\
\vsp
&\times&\left[\frac{\Gamma(6-n)f_0}{({\rm detU})^{n/2}(V-i\epsilon)^{6-n}}  \right.  \nn \\
\vsp
&+&
\left. \frac{\Gamma(5-n)f_2}{-2({\rm detU})^{n/2+1}(V-i\epsilon)^{5-n}}
\right]  \label{eq:finalform}
\eea
The numerators $f_0,f_2$ represent the coefficient of 
$\ell^0, \ell^2$ term, respectively,  after  projection operator is applied.
They are also the function of dimension $n$.
%%%%%%%%%%%%%%%%%%%%%%%%%%%%%%%
\subsection{\label{subsec:Reg} Regularization Method}
Next step is the regularization of UV- and IR-divergences.
  By adopting $n$-dimensional
regularization method,
any integrand $F$ of Feynman parameter integration is regarded as a function of  $\ve=2-n/2,~F(\ve)$. 
We adopt two methods for regularization. 
\subsubsection{ Linear Expansion method} 
First one is very simple and powerful if the accuracy of numerical integration 
is sufficiently guaranteed. We call it Linear Expansion method (LE).
Just after the dimensional regularization method was introduced [\onlinecite{n-dim},\onlinecite{n-dim2}],
 the analyticity with respect to $\ve$ was discussed extensively. 
 It is shown that the Feynman amplitude is a meromorphic function of $\ve$ 
 [\onlinecite{n-dim},\onlinecite{n-dim2},\onlinecite{Nakanishi}]. 
This is a key point to utilize the LE-method to the Feynman amplitude. 
The followings are the steps to get the divergent and finite terms.
\begenu
\item
Calculate $G(\ve) =\int F(z_j,\ve)\prod dz_j$  
for various values of $\ve=\ve(i)$. ~$(i=1,2,\cdots M)$.
\item
 We set  $\ve(i)=1/\alpha^{i+14}$ by taking relevant value $\alpha$.
\item
According to the analyticity, we can expand  $G(\ve(i))$ in Laurent series.
In our case, it is evident that the expansion starts from $(1/\ve(i))$ because of the lack of 
two-loop counter terms. We truncate the series at  O($\ve(i)^{M-2}$).
\bea
G(\ve(i)) &=& C_{-1}\frac{1}{\ve(i)}+C_0 + C_1\ve(i)+\cdots +C_{M-2}\ve(i)^{M-2} \nn \\
&=&\sum_{j=-1}^{M-2}C_j\{\ve(i)\}^j
~~~~(i=1,2,\cdots M)
\eea
The coefficients $C_{-1}$ and $C_0$ correspond to the divergent and finite parts, respectively.
In the case $\ve=2-n/2$, $C_{-1}$ expresses the UV-divergent part and 
if we set  $\ve \rar \ver =(n/2-4), C_{-1}$ represents the IR-divergent part.
\item
To get  $\{C_{j}\}$,  we multiply the inverse of  $M\times M$ matrix $A$, 
whose element is  $A(i,j)=\{\ve(i)\}^j,~(i=1,\cdots M,~j=-1,0,\cdots, M\mhs2)$, to  M-component vector $G(\ve(i))$. 
\bea
C_{j}=\sum_{i=1}^M A^{-1}(j,i) G(\ve(i)),~~(j= \mhs 1,0,\cdots, M\mhs 2)
\eea
\item
In order to improve  the convergence,  
we set  $M\sim18$ and $ \alpha \sim 1.1$ by trial and error.  Examples
setting  these parameters are  shown in 
[\onlinecite{LEM},\onlinecite{Donc224}]
\item
Various methods are known to extract $C_{-1},~C_0$ from  $G(\ve(i))$ [\onlinecite{Sidi}],
however, LE method is simple and appropriate in our case.
\endenu
In order to get the reliable value of $C_{-1},C_0$ up to 4 digits, we need the 
accuracy of the numerical integration  at least 8 digits. 
%%%%%%%% %%%%%%%%%%%%%%%%%%%%%%%
\subsubsection{Subtraction Method}
To complement the above calculation, we also  adopt the well known subtraction method to
separate divergent part and finite part.
We  extract $1/\ve$~ singularity from $G(\ve)$ when
one of Feynman parameters  approaches  0,
($x\rightarrow 0$). The followings are the steps to extract the singularity.
\begenu
\item
First we transform the Feynman parameters $(z_1,z_2,\cdots,z_6)$ into the appropriate [0,1] variables
$(x,y,u,v,w)$ depending on the topology. Key point is to factorize the function detU=$x\times z(x,\cdots)$,
where $z(0,\cdots)\ne 0$ . Singular behavior $(1/\ve)$ comes from the factor (detU)$^{n/2}$ in Eq.(\ref{IR0})
\item
 The following formula is effective to extract the factor $(1/\ve)$
 for vertex type correction.
\bea
I&=& \int_0^1 x^{\ve-1}F(x,\ve)dx= \frac{1}{\ve}F(0,0)
+\frac{\partial F(0,0)}{\partial \ve}  \nn \\
\vsp
 &+& \int_0^1\frac{F(x,0)-F(0,0)}{x} dx +O(\ve) \label{vertex-type}
\label{vertex-type}
\eea 
\item
In the case where there is self-energy type diagram, 
the factor $x^{\ve-2}$ appears in the head of integrant. 
If we expand it in $\ve$  by using analytic continuation 
the following formula is obtained.
\bea
I&=&  \int_0^1 x^{\ve-2} F(x,\ve) dx \nn\\
\vsp
&=& \frac{1}{\ve}\frac{\partial F(0,0)}{\partial x} 
+\frac{\partial F(0,0)}{\partial x}+\frac{\partial^2 F(0,0)}{\partial \ve \partial x}
-F(1,0)\nn \\
\vsp
&+&\int\nolimits_0^1 
\frac{ {\partial F(x,0)}/{\partial x} -  {\partial F(0,0)}/{\partial x}}{x}dx +O(\ve) 
\eea
\endenu
 We use this method partly to complement  the LE method.  
\subsubsection{Counter terms}
 As for counter terms, GRACE has a library of renormalization constants at one-loop level
 based on OS-renormalization scheme. 
  We make use of this library for 70-diagrams composed of  counter terms. 
Generally speaking, it is necessary  to expand one-loop  renormalization constants  up to order $\ve=2-n/2$ at 
two-loop level. 
However, the  divergent part  of diagrams composed of CT does not
contribute to $a_\mu$ ,  the $O(\ve)$ term is unnecessary in our case.  
Here we comment on the wave function renormalization constant of goldstone fields $\chi,\chi_3$.  
We keep the finite part of the constant in the form $(-1/2)\{d\Pi^{\chi\chi}(q^2)/dq^2\} ~ at~ q^2=M_W^2$. 
 However, the final answer is independent of the finite part. For  the renormalization of unphysical fields, 
 the UV-divergent part is only useful to erase divergence.
%%%%%%%%%%%%%%%%%%%%%%%%%%
\section{\label{sec:Numcal}Numerical Calculation}
%%%%%%%%%%%%%%%%%%%%%%%%%%
\subsection{\label{sec:DEmethod} Double exponential method}
%%%%%%%%%%%%%%%%%%%%%%%%%%%%%%%%%
The final step to get the value $a_\mu$  is the numerical integration over Feynman parameters.
We employ trapezoidal rule with Double Exponential (DE) transformation method  [\onlinecite{HTakahashi}]. 
It is also called as $tanh$-$sinh$ transformation method.
It is very powerful if the integrand has singular behavior at the edge of  the integration domain.
Speed of convergence is accelerated by the DE transformation,
\bea
I=\int_{0}^{1}dx f(x)\rar  x = \phi(t)=\hlf
\left\{1+\tanh\left(\frac{\pi}{2}\sinh(t)\right) \right\} \nn \\
\eea
The maximum dimension of multiple integration is five. We apply DE-method to
any integration variable involved. 
As we need  the accuracy   greater than 8 digits  to see  the UV cancellation, the
adaptive Monte Carlo method is not adopted in our two-loop calculation. 
\vspace{-0.3cm}
%%%%%%%%%%%%%%%%%%%%%%%%%%%%%%%%%%%%%%%
\subsection{\label{sec:Criterion}Criterion to ensure the validity of the result}
In order to ensure the validity of our results, we impose several conditions given below.
\begenu
\item
Well known QED two-loop value is reproduced  up to 7 digits.
\item
UV-divergence is cancelled.
\item
IR-divergence is cancelled.
\item
The result is independent of non-linear gauge parameters.
\item
In some cases (examples: topology 4,5-a,5-b,7,9-1,9b and 10) ,
we can perform loop-integrations $\ell_1$ and $\ell_2$ successively.
(We call it successive method.) We obtain the same value as
 the direct method previously shown.
\endenu
In all these cases, if we have plural methods to evaluate, we compare the numerical values 
to ascertain the validity.
We demonstrate how the conditions are cleared  by showing the examples 
 in Appendices.
%%%%%%%%%%%%%%%%%%%%%%%%%%%%%%%%%%%%
\subsubsection{Non-linear gauge (NLG) parameter independence}
Originally non-linear gauge was introduced
to reduce the number of diagrams, particularly containing boson-boson couplings
[\onlinecite{NLG-Fujikawa,NLG-Jog,NLG-Shizuya,NLG-Das,NLG-Romao,Boudjema}].
Here, we adopt NLG to check the validity of our calculation.
The gauge fixing Lagrangian is constructed as,
\bea
\mathscr{L}_{GF}=-\frac{1}{\xi_W}F^+F^- -\frac{1}{2\xi_Z}(F^Z)^2
-\frac{1}{\xi}(F^A)^2
\eea
where
\bea
F^{\pm}&=&\left( \partial^\mu\mp ie \xnla A^\mu \mp i\frac{e c_W}{s_W}\xnlb Z^\mu \right)W_\mu^\pm \nn\\
&+&\xi_W\left ( M_W\chi^\pm+\frac{e}{2s_W}\xnld H\chi^\pm 
\pm i\frac{e}{2s_W}\xnlk \chi_3\chi^\pm     \right) \nn \\
F^Z&=&\partial^\mu Z_\mu + \xi_z\left( M_Z \chi_3 + \frac{e}{2s_Wc_W}\xnle H\chi_3\right) \nn \\
F^A&=&\partial^\mu A_\mu
\eea
Here, $\xnla,\xnlb,\xnld,\xnle$ and $\xnlk$ are non-linear gauge parameters specific to 
this formalism. The parameters $s_W$ and $c_W$ are the $sine$ and $cosine$  of Weinberg angle $\theta_W$. 
In our calculation we set $\xi=\xi_W=\xi_Z=1$ to make the gauge boson propagators 
simple.
NLG parameters are distributed among so many diagrams of different types of topologies.
So it is very powerful if we can verify the cancellation of these NLG parameters.
We show  the sample of cancellation in  Appendix \ref{appNLG}.
%%%%%%%%%%%%%%%%%%%%
\subsubsection{Successive method}
Diagrams with self energy type two-point function can be 
calculated by successive method using renormalized two point
function. An example is diagram with 
$(\gamma-\gamma)$ or $(\gamma-Z)$ vacuum polarization type diagrams.
We decompose the renormalization constants $\delA,\delZA,\delAZ,\delta M_Z^2 $ etc.
 into components according to the particles
involved in the loop.
 By adding the  counter term
 to corresponding one-loop unrenormalized two-point function, 
 one-loop ($\ell_1$) integration is performed without divergence and
we obtain the renormalized two-point function $\Pi_R$.
By inserting $\Pi_R$
into the second loop($\ell_2$),  we get
finite value of  $a_\mu$ . We use this alternative method to 
reconfirm the results obtained  by the methods given in section \ref{sec:Logic}.
An example is shown in Appendix \ref{appSM}.
%%%%%%%%%%%%%%%%%%%%%%%%%%%%%
\section{\label{sec:Result}Results of our calculation}
%%%%%%%%%%%%%%%%%%%%%%%%%%%%%
As the physical input parameters, 
we use the following fermion and boson masses (unit GeV).\\
$m_{\mu}= 105.6583745\times 10^{-3}$,     
 $m_{e} = 0.5109989461\times 10^{-3}$, $m_{\tau}= 1.77686$
  $m_u = 0.3$, $m_c = 1.5$,
  $m_t = 173.1$,$m_d = 0.3$,
  $m_s = 0.5$, 
 $m_b=4.18$, $M_W = 80.385$,
  $M_Z= 91.1876$, $M_H=125.09 $. 
   We also choose the fine structure constant  in the Thomson limit,
     $\alpha$ =1/137.035999139.  
     
  After clearing all the conditions given in section \ref{sec:Criterion} we get 
  the two-loop ELWK corrections to $a_\mu^{EW:OS}$[2-loop] in terms of $(\alpha,M_Z,M_W,M_H,m_f)$. 
  The loop expansion is done by using 
  $\alpha, \alpha^2$, successively. 
  Among 1780 diagrams, we exclude 9 pure QED diagrams consisting of only $(e,\mu,\tau,\gamma)$ and
  6  diagrams  containing vacuum polarization composed of quark loop.   
  Then the final result becomes,
   \bea
  a_\mu^{EW:OS}[2\mhs{\rm loop}] = (-36.76 \pm 0.3)\times 10^{-11}.
  \eea
 The errors in the above and the following expressions are  limited to 
  the numerical integration error and the uncertainty of  input  parameters $M_W, M_H,M_Z, m_t,m_b$. 
 The masses of light quarks are fixed in our model.   
 
 We show the fermionic and bosonic part of two loop correction separately for reference.
\bea
a_\mu^{EW:OS}[2\mhs{\rm loop}]_{\rm fermion}&=&-18.34(\pm 0.2) \\
\vsp
a_\mu^{EW:OS}[2\mhs{\rm loop}]_{\rm boson}&=&-18.42(\pm 0.1)
\eea  
As we mentioned before, we adopt OS renormalization, however, 
the expression in the preceding works is parametrized using Fermi constant
$ G_F=1.1663787 \times 10^{-5} ~{\rm GeV}^{-2}$ and  $\alpha$ \cite{Weak2}. 

The difference of the 2-loop correction between our value and 
that in $G_F$ parametrization is due to the fact that 
one loop correction in $G_F$ parametrization
partially includes the $\alpha^2$ correction in our scheme. 

So the comparison should be done to the sum of one- and two-loop. 
The one loop correction in our OS scheme is written down as follows.
\bea
&& a_\mu^{EW:OS}[1\mhs {\rm loop}]= \frac{\alpha m^2_\mu \mz}{16\pi(\mz-\mw)\mw} \times \nn \\
\vsp
&&\left[ \frac{5}{3}+ \frac{1}{3}\left( \frac{4\mw}{\mz}-3\right)^2
+\mathscr{O}\left(\frac{m^2_\mu}{\mw}\right)+ \mathscr{O}\left(\frac{m^2_\mu}{M_H^2}\right)\right] \nn \\
\eea
 We can carry out the numerical calculation without any approximation and get  the value.
\bea
a_\mu^{EW:OS}[1\mhs {\rm loop}]= 187.99(\pm 0.2) \times 10^{-11} 
\eea
By summing up one and two loop weak corrections, our result is as follows.
 \bea
  a_\mu^{EW:OS}[1{\rm+}2\mhs {\rm loop}] = 151.2( \pm 1.0)
   \times 10^{-11} \label{eq:onetwo} 
 \eea 
 Here we add  the error due to 
 neglecting  the uncertainty in electroweak loops involving hadrons.
 
 When we compare our result with the value obtained by
using $G_F$ parametrization, we need the naive free light quark model calculation with 
the same quark masses as ours.
This is given\footnote{
 In ref.\cite{Marciano3}, the contribution of light quarks in FQM is 
 $a_\mu^{EW(2)}(e,\mu,u,c,d,s)=-(4.0+4.65)\times 10^{-11}=-8.65\times 10^{-11}$}
 in ref.\cite{Marciano3}. 
 In this case the two loop correction becomes, $-42.97(\pm 1)\times 10^{-11}$.
 In the $G_F$ parametrization, the one loop correction becomes  $194.80(\pm 0.01)\times 10^{-11}$\cite{Weak2},
 so that we get,
\bea
 a_\mu^{EW}[1{\rm+}2\mhs{\rm loop}]_{\rm FQM} = 151.8( \pm1)
   \times 10^{-11}.~~~~
 \eea

This is consistent with our value Eq.(\ref{eq:onetwo})

We also add a comment on the relation between the 
well known PDG value \cite{PDG}
shown below and our value.
If we include the hadronic correction to light quark contribution,
by  adding  the difference of the following expression \cite{Weak2},
 \bea
a_\mu^{EW(2)}(e,\mu,u,c,d,s)=(-6.91\pm0.20\pm0.3)\times10^{-11} \nn\\
\eea
and the value quoted in the footnote below\cite{Marciano3},
our value becomes as follows.
\bea
 a_\mu^{EW}[1{\rm+}2\mhs{\rm loop}] = 152.9( \pm1) \times 10^{-11}
\eea
It is in agreement  with the following PDG value \cite{PDG} within errors.
\bea
a_\mu^{EW} = 153.6(\pm1)\times 10^{-11} \label{eq:pdgonetwo}
\eea
%%%%%%%%%%%%%%%%%%%%%%%%%%%%
\section{\label{sec:DC}Discussions and Comments}
%%%%%%%%%%%%%%%%%%%%%%%%%%%%
We developed the system to calculate the full ELWK two-loop corrections to $a_\mu$ ,
by fully using GRACE and FORM  on the basis of OS-scheme.  
 The work we need beforehand is only to prepare several files
 which only depend on the type of the topology of diagrams
 as we explained in section I\ref{sec:Outline}.
We adopt the dimensional regularization to regularize UV- and IR-divergences
and to get finite gauge invariant values of the physical quantity. 
To extract the $(1/\ve)$ terms, we use Linear Expansion Method explained 
in sectionIII-B. This method is very simple and attractive, compared
with the conventional method to take out the $(1/\ve)$
term by extrapolating one of the Feynman parameters close to 0 .
If we adopt the conventional method, it is crucial to introduce 
the most suitable transformations from $(z_j)$ to [0,1] integration variables  $(x,y,u,v,w)$.
Furthermore, we need rather complex operations including differentiation of the amplitude, etc.
 As a result, the necessary CPU-time increases extensively.

In the case of Linear Expansion method (LE), however, 
the choice of  integration variables is not sensitive to get the reliable results and
this method  decreases the number of operation drastically. 
It is sufficient to define the quantity as function of $\ve(=2-n/2)$ .
We only need to treat Dirac matrices and various vectors appeared in the 
numerator, in $n$-dimension. 
This is easily done by using symbolic manipulation system such as FORM.
The operation is simple and we can make use of the resultant short sources 
 for both UV-($\ve>0$) and IR-($\ver=-\ve > 0$) regularization and 
 also to get finite results.
We conclude that LE-method is 
the most simple and reliable method, at this moment.
In order to get  reliable physical value by this method, 
high precision numerical integration over Feynman parameters is 
inevitable.  The DE-method introduced in section \ref{sec:DEmethod} is the suitable candidate.

 Introduction of NLG-parameters makes the calculation very complex, however, 
 it is very powerful to check the calculation of so called Boson contribution.
 The number of diagrams consisting of different types of topology 
 are connected  through NLG-parameters. The maximum number of 
 diagrams mutually entangled  reaches  864. So this is a very tough 
 condition to be cleared. 
 
 By making use of these technical approaches mentioned above, we clear 
all the constraints given in section \ref{sec:Criterion} .
Namely, (i) reproduction of QED values, (ii)(iii) cancellation of UV-and IR-divergences, 
 (iv) Independence of NLG-gauge parameters.
 We show some samples in Appendices how they are cleared. 

 The final value of the sum of one and two loop weak corrections is approximately the same
 as  the one obtained by previous works using different parametrization.
 
 Based on this work we can proceed to construct 
 PNQFT(Perturbative Numerical Quantum Field Theory), which we discussed 
 in section \ref{sec:PNQFT}.   Wide range of application to ELWK higher 
 loop expansion for several physical reactions will be opened.  
 We expect that this work provides the fruitful foundation to formulate PNQFT.
%%%%%%%%%%%%%%%%%
\begin{acknowledgements}
%%%%%%%%%%%%%%%%%
We would like to thank Prof.T.Kaneko for his important contribution to construct the 
framework of  calculation at the early stage of this work. We also wish to 
thank Prof.K.Kato, Prof.F.Yuasa and Prof.M.Kuroda  for discussions. Last but not least, we 
express our deep appreciation to late Prof.Y.Shimizu for his continual encouragement.
This research is partially supported by Grant-in-Aid for 
Scientific Research (15H03668,15H03602) of JSPS and
 Grand-in-Aid for High Performance Computing with General Purpose Computers
(Research and development in the next-generation area) of MEXT.

\end{acknowledgements}
%%%%%%%%%%%%%%%%%%%%
\appendix
%%%%%%%%%%%%%%%%%%%%%%%%%
\section{Reproduction of QED two-loop value}
%%%%%%%%%%%%%%%%%%%%%%%%%
QED two-loop value is reproduced correctly. 
\begin{center}
\begin{tabular}{lcc}
  &   &Unit =$(\alpha/\pi)^2 $\\
Analytic expression&　　&-0.328478996 \\ 
0ur value & & -0.328479821  
\end{tabular}
\end{center}
%%%%%%%%%%%%%%%%%%%%%%%%%%%%%%%%%%%%
\section{Classification of Diagrams to Check Numerical Values}
\begin{figure}[H]
\begin{minipage}{\columnwidth}
\centering
\includegraphics[bb= 0 0 828 314,scale=0.25]{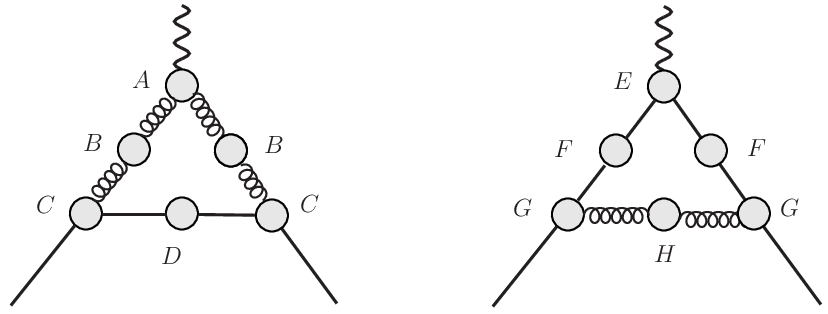}
\end{minipage}
\caption{\label{fig:topfb} Classification of Diagrams}
\end{figure}

%%%%%%%%%%%%%%%%%%%%%%%%%%%%%%%%%%%%
In Fig.\ref{fig:topol}, we show types of topology to
 formulate the two-loop contributions.  
 However, in order to check the consistency 
 of numerical values, it is useful to classify diagrams by distinguishing 
 fermion and boson lines in each diagrams in Fig.\ref{fig:topol}.
 We briefly figure out the classification method in Fig.\ref{fig:topfb}.
%%%%%%% Figure : Fig-2 %%%%%%%%%%%%%
 In the figure, the straight lines and wavy lines represent 
fermion and boson, respectively. The circle indicates 
one-loop diagram. We classify diagrams depending on the place where 
one-loop diagram is inserted. 
It is summarized in the following Table \ref{tab:Class8}.
You can easily see which one of Fig.\ref{fig:topol}  is classified 
into which category.
%%%%%%%%%%  Table  I  %%%%%%%%%%%%%
 \begin{table}[H]
\centering
\caption{\label{tab:Class8}{
 Classification of two loop diagrams. 
Depending on the position (A$\sim$H) of one loop diagram or one loop 
counter term, we give names shown below. For example, LAD-I and LAD-II 
correspond to topology (3) in Fig.1. The typical diagram belonging to
LAD-I is fermion triangle with $\gamma\mhs\gamma\mhs Z$ legs. }
}
\begin{ruledtabular}
\begin{tabular*}{80mm}{|c|c|c|c|} 
A $\rar$ LAD-I & B $\rar$ SLF-I& C$\rar$ VTX-1& D$\rar$VCP-I \\
\hline
E$\rar$ LAD-II&F$\rar$ SLF-II & G$\rar$ VTX-II&H$\rar$ VCP-II \\
\hline 
\multicolumn{2}{|c|}{{Fig.1(1) } $\rar$ {CRL}}
&\multicolumn{2}{c|}{{Fig.1(6) } $\rar$ {DBT} }\\
\end{tabular*}
\end{ruledtabular}
\end{table}
In the above Table \ref{tab:Class8} we add two types of topology having  no divergence, 
namely, Fig.\ref{fig:topol}-(1) and (6).
%%%%%%%%%%%%%%%%%%%%%%%%%%%
\section{UV-cancellation}\label{appUV}
\begin{figure}[h] 
\begin{minipage}{\columnwidth}
\centering
\includegraphics[scale=0.3,bb=0 0 618 428]{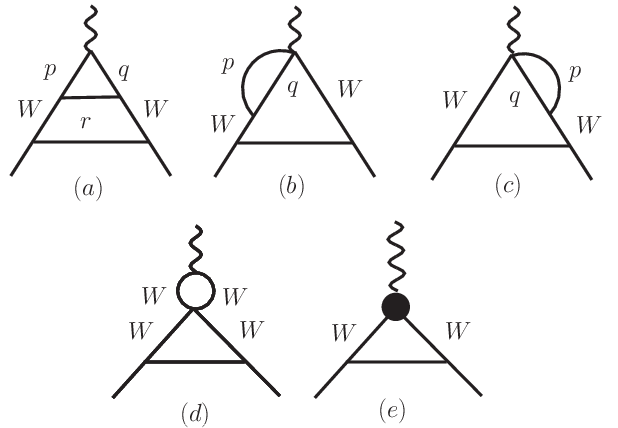}
\end{minipage}
\caption{\label{fig:UVcancel}Sample diagrams with $W$}
\end{figure}

%%%%%%%%%%%%%%%%%%%%%%%%%%%
In this Appendix, we show the cancellation of UV-divergence
in linear gauge ('t\,Hooft-Feynman gauge). 
Examples of a group of diagrams are shown in Fig.\ref{fig:UVcancel}.
The diagrams in Fig.\ref{fig:UVcancel} belong to several groups in Table \ref{tab:Class8}.
%%%%%%%%% Figure : Fig-3 %%%%%%%%%%%%%%%%%

In this case, total 13-diagrams and 1-counter term (e)
make a group to cancel UV-divergence. 
In the Table \ref{tab:UVFG}, the coefficient of 
$C_{UV}^{(2)}=(1/\ve-2\gamma+2\ln(4\pi))$, corresponding to 
each diagram is shown.
As you can read from  the Table \ref{tab:UVFG},
the cancellation is marvelous,  up to almost 15 digits. \\
%%%%%%%%%%  Table  II   %%%%%%%%%%%%%%%%%%%
\begin{table}
\centering
\caption{\label{tab:UVFG}
 Sample of UV cancellation in 't\,Hooft-Feynman gauge.
UV cancellation is confirmed by summing up the coefficient of 
$C_{UV}^{(2)}$ of a set of diagrams. Here we show  the set given in 
Fig.3 as a sample. The 15-digit cancellation is realized in the sum.
}
\begin{ruledtabular}
\begin{tabular*}{\columnwidth}{@{\extracolsep{\fill}}llll@{}}
Diagram & Particles  on &   Value ~~$( \rm{unit~ }10^{-11})$\\
~~in  Fig.3  &  ~~line $(p,q,r)$ &  &  \\
\hline
(a)& $W-W-Z$& ~1.72758038865755734 \\
(a)& $W-W-\gamma$ & ~0.49552373441415571 \\
(a)& $\chi-\chi-H$ & ~0.05700266982235161 \\
(a)& $\chi-\chi-\chi_3$& ~0.05700266982235161\\
(a)& $c^--c^--c^Z$&  -0.02214846652125073\\
(a)&$c^--c^--c^\gamma$& -0.00635286838992507\\
(a)& $c^+-c^+-c^Z$& -0.02214846652125073\\
(a)& $c^+-c^+-c^\gamma$ & -0.00635286838992507\\
(b)&$W-Z$ & -0.17085951453900741 \\
(b)&$W-\gamma$ & -0.04900781767402955\\
(c)&$W-Z$ & -0.17085951453900741\\
(c)&$W-\gamma$ &  -0.04900781767402955\\
(d)& &  -2.63840950598091337\\
(e)& CT ($\gamma-W-W$) & ~0.79803737751292251 \\
\hline
 & Sum &  -0.00000000000000012\\
\end{tabular*}
 \end{ruledtabular}
\end{table}
%%%%%%%%%%%%%%%%%%%%%%%%%
\section{ IR-cancellation}\label{appIR}
\begin{figure} [H]
\begin{minipage}{\columnwidth}
\centering
\includegraphics[scale=0.30,bb=0 0 805 502]{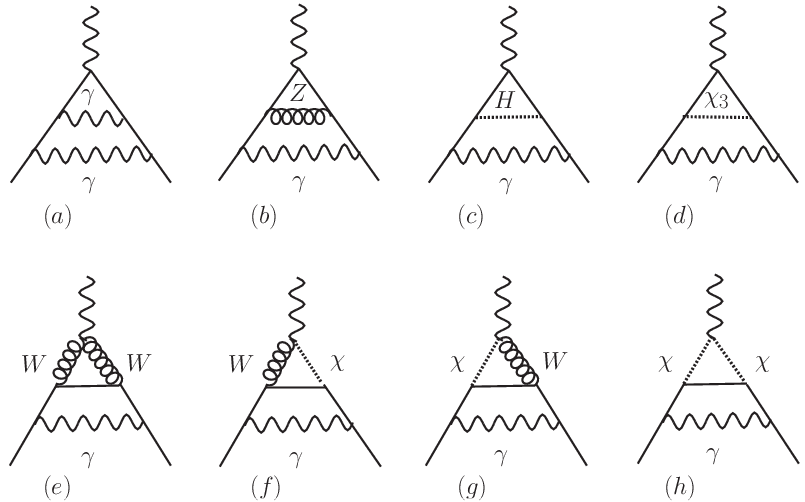}
\caption{\label{delW} 
All the two-loop diagrams containing  IR-divergence
}
\end{minipage}
\end{figure}

%%%%%%%%%%%%%%%%%%%%%%%%%
LE method is applied to check IR-cancellation.
Among two-loop diagrams,  the 8 diagrams in Fig.(\ref{delW}) have IR-divergence.
The diagrams with CT also have IR-divergence through $\delW$ (19 diagrams) and 
$\delta Z_\mu^{1/2}$ (28-diagrams).
%%%%%%%%%%%%%%%%%%%%% 
The IR-divergence at two-loop level is proportional to 
$C_{IR}^{(2)}= (-1/\ver -2\gamma+2\ln(4\pi))$,  $\ver=(n/2-2)>0$. 
It is easily shown that the IR-divergence coming  from $\delW$ cancels 
among the 19 CT-diagrams. As for the diagrams with $\delta Z_\mu^{1/2}$,
IR-divergence is  cancelled by the corresponding two-loop diagrams. 
 We show  coefficients  of $C_{IR}^{(2)} $ in Table \ref{tab:IRcan}.
 The correspondence between small photon mass ($\lambda$) method and LE-method for IR-regularization
is checked in the case of QED ladder diagram.
Analytic value of a coefficient of  $\ln(\lambda^2/m_\mu^2)$
in unit of $(\alpha/\pi)^2$ is $(1/4)$  \cite{Karplus}. 
It is  0.249999998 by our calculation 
using small photon mass. 
In LE-method, the  coefficient of $C_{IR}^{(2)}$ becomes  0.249999999. 
 We understand 
the correspondence between  $\ln(\lambda^2/m_\mu^2)$ and  $C_{IR}^{(2)}$  is established.
As we  show in Table \ref{tab:IRcan},  no IR-divergence remains in the final expression.
%%%%%%%% Figure : Fig.4   %%%%%%%%%%%%%
%%%%%%%% Figure :Fig.5   IR-through Renormalization Constant %%%%%
\begin{figure} [H] 
\centering
\includegraphics[scale=0.32,bb=0 0 545 237]{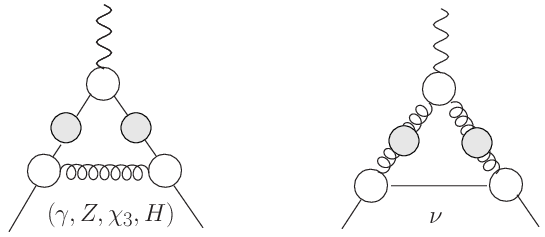}
\caption{\label{fig:808line} 
Diagrams containing IR-divergence through $\delta Z$.
The straight lines and wavy lines represent 
fermion and boson, respectively. 
The circle represent $\delta Z_\mu^{1/2} $ or $\delta Z_W^{1/2}$.
In the case where  the corner includes  $\nu$-particle, both of $\delta Z^{1/2}_{\mu}, \delta Z^{1/2}_W$ are taken into account.
The gray circles and white circles  correspond to Self CT and Vertex CT in 
Table \ref{tab:IRcan} , respectively.
 }
\end{figure}
%%%%%%%%%%%  Table  III  %%%%%%%%%%
\begin{table*}
\caption{\label{tab:IRcan}
Cancellation of IR-divergence is shown. IR cancellation is confirmed 
by summing up the coefficient of $C_{IR}^{(2)}$. All the diagrams having 
IR-divergence are shown in FIG.4. Typical diagrams having IR-divergence 
through the renormalization constants are shown in FIG.5.
}
\begin{ruledtabular}
\begin{tabular*}{\textwidth}{@{\extracolsep{\fill}}lcccc@{}}
Type of correction &Numerical Value (unit $10^{-11}$) &&  \\
\hline
Diagram Type & $\delta Z_\mu^{1/2}$ Vertex CT &  $\delta Z_\mu^{1/2}$ Self CT & two-loop diagram &sum  \\
\hline
 Neutral Type  &    &   &  diagram  &  \\
    $\gamma$  &  $-134887.2756 \times 3$ & $134887.2756 \times 2$  &(a)   $+134887.2755$ & $-1.6 \times 10^{-4}$ \\
    $ Z-boson$   &$ 0.217294 \times 3  
    $ & $ -0.217294 \times 2 $ &(b) $-0.217294$ & $ -2.0 \times 10^{-9}$  \\
 $ Higgs, \chi_3$ & $1.67970\times 10^{-6} \times 3$ & $-1.67970 \times 10^{-6} \times 2$ & 
$(c,d)-1.67970 \times 10^{-6}$  & $ 2.6 \times 10^{-13}$ \\
\hline
 Charged type &      &    &   & \\
   $ W^{\pm} ,\chi^{\pm}$ & -0.435625& $----$ &(e)$\sim$(h) 0.435625    & $+3.4\times 10^{-12}$\\
\end{tabular*}
\end{ruledtabular}
\end{table*}

%%%%%%%%%%%%%%%%%%%%%%%%%%%%%%%%%%%%%%%%
\section{\label{app:808sample}Sample calculation of the two-loop diagram}
%%%%%%%%%%%%%%%%%%%%%%%%%%%%%%%%%%%%%%%%
As a sample, we show the calculation of the 
two-loop diagram which contains  both UV- and IR-divergences.
The diagram is shown in Fig.\ref{delW}-(b). The  line numbers
are given in Fig.\ref{fig:808line}.
Following  Eq.(\ref{eq:finalform}) given in section III-A,
the essential part of expression of two-loop diagram contribution  is written as the following form.
\bea
F_0&=&\int \prod dz_j 
\left[\frac{1}{({\rm detU})^{-3\ve}}\frac{f_0}{(\mathscr{D}-i\epsilon)^{2+2\ve}} \right] \label{eq:E1}
\eea
\bea
F_2&=&\int \prod dz_j 
\left[\left(-\hlf\right) \frac{1}{({\rm detU})^{2-3\ve}}\frac{f_2}
{(\mathscr{D}-i\epsilon)^{1+2\ve}} 
\right]  \label{eq:E2} \nn \\
\eea
Quantities in Eq.(\ref{eq:E1}),(\ref{eq:E2}) are expressed by  Feynman parameters 
 $z_j,~(j=1,\cdots,6)$, ($\sum_jz_j=1$). Masses are made dimensionless
 using muon mass.
\bea
&&V =-z_5+z_5^2(z_{23}+z_6)/{\rm detU} + z_{1234} + z_6\mz  \nn\\
\vsp
&&{\rm detU} =z_6z_{12345} +z_3z_{45}+z_2z_{145}+ z_1z_3 \nn  \\
\vsp
&&\mathscr{D}\equiv {\rm detU}\times V =
(z_{14}+z_{23})z_{14}z_{23} + z_{23}^2z_5  \nn \\
\vsp
&&~~~~+ (z_{14}+z_{23})^2z_6 + (z_6\mz){\rm detU} 
\eea
where $z_{ij\cdots k}=z_i+z_j+\cdots + z_k$.
Functions $f_0, f_2$ are expanded in $\ve$ up to $O(\ve)$.
\bea
f_0=f_{00}+\ve f_{01},~f_2=f_{20}+\ve f_{21},~\ve=(2-n/2)~~~
\eea
 The $f_0$ and $f_2$ part give IR- and UV-divergences, respectively.
 By adopting  DE and LE methods for Feynman parameter integration,
 we get the following expansion. ($\ver=-\ve$)
  \bea
 F_0&\rar&\left( \frac{C_{-1}}{-\ver} + C_0 + C_1(-\ver) +\cdots\right) \nn \\
 \vsp
 F_2&\rar&\left( \frac{D_{-1}}{\ve} + D_0 + D_1\ve +\cdots\right) \nn 
 \eea 
  We multiply the $\Gamma$ functions arising from n-dimensional 
 integration and factor $1/(4\pi)^n$ , to the above quantities,
 \bea
 \frac{\Gamma(6-n)}{(4\pi)^n}F_0&\rar &
 \left[C_{-1}C_{IR}^{(2)} +2C_{-1}+C_0+O(\ver)\right]  \nn \\
 \vsp
 \frac{\Gamma(5-n)}{(4\pi)^n}F_2&\rar&
 \left[D_{-1}C_{UV}^{(2)}+D_0+O(\ve)\right]  
 \eea
 
 We must pay attention that the term $(2C_{-1})$ appears,
 because $\Gamma(6-n)$
 contains the term $(1-2\ver)$.
 In order to get the correct finite value, we must calculate
 the counter terms using the same regularization method for 
 both IR- and UV-divergences as in two-loop case. In this case, we need 
 muon wave function renormalization constant  $\delta Z_\mu$.
The $\delta Z_\mu$ is obtained by calculating muon self energy diagrams.
The $\delta Z_\mu(\gamma)$ represents photon exchange part and  
$\delta Z_\mu(Z)$ represents Z-exchange part.
 %%%%%%%%%    Figure 6 : Fig.6  diagram with both  UV and IR %%%%
 \begin{figure} [H] 
\centering
\includegraphics[scale=0.32,bb=0 0 723 252]{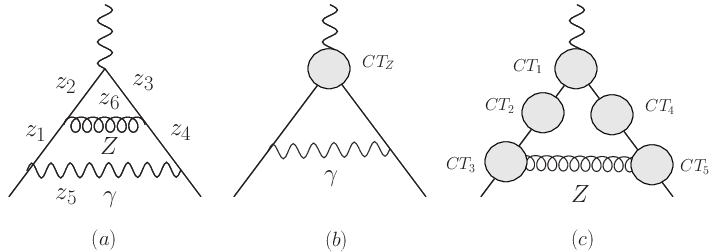}
\caption{\label{fig:808line}
 Diagram containing both UV- and IR-divergences.
 $CT_Z,CT_1 \sim CT_5$ represent the CT-terms.
 }
\end{figure}

In Fig.\ref{fig:808line}-(b), $CT_Z$ represents $\delta Z_\mu(Z)$,
and in Fig.\ref{fig:808line}-(c), $CT_1 \sim CT_5 $represent only 
IR-divergent part of $\delta Z_\mu(\gamma)$. 
These IR-divergent part cancels that of Fig.\ref{fig:808line}-(a)
as is shown in the Table \ref{tab:IRcan}.
   We will show the numerical results to see the situation clearly. 
 (unit=$10^{-11}$)
 \bea
C_{-1} &=&  -0.217294 \nn \\
C_{0}&=& ~~1.672727\nn \\
D_{-1}&=&  ~~24621.375584 \nn \\
D_0&=&  -246732.662539 
 \eea
The term $CT_Z$ has both UV and finite parts.
\bea
(CT_Z)_{UV}&=& -24621.375584 \nn \\
(CT_Z)_{f}&=& ~~246728.863173
\eea   
UV-cancellation is excellent and cancels 17 digits.
\bea
D_{-1} +(CT_Z)_{UV}= 3.84\times 10^{-12}
\eea 
To cancel IR part of  Fig.\ref{fig:808line}-(a) ,
 we adopt IR-part of  $\delta Z_\mu(\gamma),CT_1 \sim CT_5$. 
 It is explained in the Table \ref{tab:IRcan}, row ($Z ~~boson$). 
The finite value is obtained as follows.
\bea
C_{0}+2C_{-1}+D_{0}+(CT_Z)_{f}=-2.561228   \label{eq:ohfinite}
\eea 
We can also regularize the IR divergent part by employing the
small photon mass $\lambda^2$, to see $\log (\lambda^2)$ 
term. As we notice that we must use the same regularization
method for both two-loop and CT diagrams. At first sight, the finite part 
changes compared with dimensional regularization method, 
however, the sum of two-loop and CT contribution is the same
in both regularization methods. We confirm it by numerical calculation.\\
As we can see from Eq.(\ref{eq:ohfinite}), we need very careful treatment
to  discuss order of  (a few) $\times 10^{-11}$ quantity of $a_\mu$.
\\
~~~~~~~~~~~
\section{NLG parameter cancellation }\label{appNLG}
As an example of  cancellation of NLG parameters, 
we show the UV-divergent part and finite part  
of diagrams which contain $\xnla^n~(n=1,2,3)$ terms.
Total number of diagrams depending on $\xnla$ amount to 864. 
In order to see how the NLG-parameters are cancelled,
we classify diagrams according to Fig.\ref{fig:topfb} and Table \ref{tab:Class8}.
We summarize the result of numerical calculation in Table \ref{tab:uvnlgc}.
In the last row in Table \ref{tab:uvnlgc}, we show the maximum 
absolute value among the individual terms in the column and its type of 
diagram, to indicate the degree of the cancellation.
First we see the cancellation of the parameter $\xnla^{n}$ in
the UV-divergent part, namely, the coefficients of $C_{UV}^{(2)}$.
It is shown in the left half part of the Table \ref{tab:uvnlgc}.
 We can see the cancellation works very well and 
it strongly guarantees the validity of our numerical calculation.
In the  right half of Table \ref{tab:uvnlgc}, we also show the finite contribution to $a_\mu$.
The column $\xnla^1 \sim \xnla^3$ show that how the NLG-cancellation works well
also in finite part. 
%%%%%%%%%%%%%%%%%%%%%%%%%%%
\section{ Successive method}\label{appSM}
In some cases, in order to check our calculation, we perform two-loop integrations successively.
By integrating the first loop ($\ell_1$), we make the renormalized effective function and 
insert it to the second loop($\ell_2$) integration. 
As an example, we calculate  the diagrams consisting of photon-Z meson 
mixing vacuum polarization. 
There are two types of topology 4 and 10 in Fig.\ref{fig:topol}.\\
In general, unrenormalized two point $\gamma-Z$ function is written as follows.
\bea
\Pi(q^2)^{un}_{\mu\nu}=\Pi^{un}_T(q^2)\left (\gmn -\frac{\qmn}{q^2}\right)
+ \Pi^{un}_L(q^2) \frac{\qmn}{q^2}   \label{vecvec} \nn \\
\eea
Each one-loop diagram contributes to $\Pi^{un}_T(q^2),\Pi^{un}_L(q^2)$ 
in the following form. 
\bea
\Pi_T^{un}(q^2)=a(q^2),~~\Pi_L^{un}(q^2)= a(q^2)+q^2 b(q^2)
\eea
\\
where
\bea
a(q^2)&=& \int\ndim \left[
\frac{A_1\ell^2+(A_2q^2+A_3)}{(\ell^2-D_Q)^2 }+\frac{A_4}{(\ell^2-m^2)}
\right]  \nn  \\
\vsp
b(q^2)&=&\int\ndim \frac{B}{(\ell^2-D_Q)^2} 
\eea
The variables $ A_1 \sim A_4,B, D_Q,m$ depend on the specific one-loop diagram.
The $A_4$-term comes from the 4-point boson coupling diagram. However,
this term drops out by the renormalization process. \\
In the followings, 
the renormalized quantity $\Pi$ are expressed as  $\hat{\Pi}$.
\bea
\hat{ \Pi}_T(q^2)&=&(\mz-q^2)\delZA-q^2 \delAZ+\Pi_T^{un}(q^2) \nn  \\
\vssp
 \hat{\Pi}_L(q^2)&=& \mz\delZA +\Pi_L^{un}(q^2) \nn \\
 \vssp
 \hat{\Pi}_T(q^2&=&0)=0,~~~~\hat{\Pi}_T(q^2=\mz)= 0
 \eea
 We notice that the renormalization conditions are fixed by $\Pi^{un}_T(q^2)$, 
 there is no freedom to renormalize  $\Pi_L^{un}(q^2)$.
By using these renormalization conditions, we can write down renormalized 
$a_R(q^2)$. It has no divergence, we
can perform loop($\ell_1$) integration.
%%%%%%%%% Table  IV    %%%%%%
\begin{table*}
%\begin{table}[H]
\caption{\label{tab:uvnlgc}
Cancellation of NLG $\xnla$ parameter is shown. Numbers are in unit of $10^{-11}$.
All the two-loop diagrams and CT terms are classified according to the TABLE II. Numbers
represent the coefficients of $\xnla,\xnla^2,\xnla^3$ after summing up the contributions of 
both two-loop diagrams and CT terms. 
The last row shows the maximum absolute value within the same column and its topology type,
 to show how the cancellation works well.
}
\begin{ruledtabular}
\begin{tabular*}{\textwidth}{@{\extracolsep{\fill}}c|r|r|rrr|rrr@{}}
Type of  &\multicolumn{2}{c|}{ Number of} & \multicolumn{3}{c|}{Coefficient of $C_{UV}^{(2)}$ (unit ~$10^{-11}$) } &
\multicolumn{3}{c}{ Finite contribution to (g-2)  (unit ~$10^{-11}$)}   \\
Diagrams & \multicolumn{2}{c|}{Diagrms} &\multicolumn{1}{c} {$\xnla$} &
\multicolumn{1} {c}{$\xnla^2$} & \multicolumn{1}{c|}{$\xnla^3$} 
&\multicolumn{1}{c} {$\xnla$} &
\multicolumn{1} {c}{$\xnla^2$} & \multicolumn{1}{c}{$\xnla^3$} 
\\
  &{ 2loop} &{CT} & &  &  &  &  &  \\
 \hline
LAD-I & 400&10 &
$1.3 \times 10^{-11}$ & -0.25876 &-0.03235 &
6.11086 & 3.98570 & 0.83255 \\
LAD-II & 468 & 6 &
$-1.0 \times 10^{-11}$ & & &
-0.12093 & & \\
VTX-I & 80 & 8 &
$ 1.1 \times 10^{-16}$& $-5.7 \times 10^{-16}$ & &
1.16937 & 2.07396 & \\
VTX-II & 90 & 8 &
$-4.7 \times 10^{-11}$&  & &
-0.83645 & & \\
SLF-I & 312 & 16 &
$3.6 \times 10^{-16}$ & 0.25876 & 0.03235 &
-3.48758 & -3.87681 & -0.83255 \\
SLF-II & 48 & 8 &
 & & &
 & & \\
VCP-I & 12 & 4 &
$ -1.1 \times 10^{-10}$& $-1.1 \times 10^{-10}$ & &
 & & \\
VCP-II & 280 & 10 &
 & & &
-0.05656 & 0.00074 & \\
CRL & 72 & 0 &
 & & &
-2.77731 & 1.18564 & \\
DBT & 18& 0 &
 & & &
 & -3.36911 & \\
\hline \hline
SUM & 1780&70 & 
$ -1.65 \times 10^{-10}$& $-1.42 \times 10^{-10}$ & 
$ < 1 \times 10^{-30}$&
0.00138 & 0.00012 & $1.1 \times 10^{-12}$ \\ 
\hline \hline
\multicolumn{3}{c|}
{~~Type of diagram} 
 &
VTX-II & LAD-I & SLF-I & 
VTX-I & LAD-I & SLF-I \\
 \multicolumn{3}{c|}{Max. absolute value} &
$-298325.54724$ &
$-8.32905$ &
$-0.19407$ &
$2763874.32283$ & $-46.38364$ & $4.41044$ \\
\end{tabular*}
\end{ruledtabular}
{\footnotesize (*) Each contribution  is calculated in quadruple precision 
method and has more effective digit than shown in the table.} 
\end{table*}
%\end{table}
%%%%%%%%%%%%%%%%%%%%%
 The result is given as Eq.(\ref{eq:renorma1}),(\ref{eq:renorma2}). 
 where  $M$ represents the mass of a particle circulating the loop.
As for $\hat{b}(q^2)$, we must check whether it is really finite or not. 
The $C_{UV}$-part of $\hat{b}(q^2)$ disappears after summing all the 
one-loop diagrams and integration of Feynman parameter $x$.
\begin{widetext}
 \bea
\hat{a}(q^2)&=&\hat{\Pi}^{AZ}_T(q^2)= \psix\int_0^1dx\int_0^1dz  
\left[A_1 \{q^2(q^2-\mz )\}[x(1-x)^2]
\frac{-2}{-zx(1-x)}\frac{1}{(q^2-M_3^2)}\right.    \nn \\
\vsp
&+&A_2\{q^2(q^2-\mz)\}[x(1-x)]\frac{1}{-zx(1-x)}\frac{1}{(q^2-M_2^2)} 
+A_3\{q^2\}[x(1-x)]\frac{1}{-zx(1-x)}\frac{1}{(q^2-M_1^2)}  \nn \\
\vsp
&+&A_3 \{q^2\}[x(1-x)]\left.\frac{-1}{V_{WZ} } \right] \label{amaster} \label{eq:renorma1}\\
\vsp
M_1^2&=&\frac{\m}{zx(1-x)},~~M_2^2=\frac{\m}{zx(1-x)} -\frac{1-z}{z}\mz  
M_3^2=\frac{\m}{zx(1-x)}-\frac{y}{z}\mz,~~V_{WZ}=\m-zx(1-x)\mz \nn  \label{eq:renorma2}\\
\eea
\end{widetext}
We also notice that the $\hat{b}(q^2)$-part  does not contribute to $a_\mu$, 
 because it is proportional to $\qmn$. 

Next step is to insert the renormalized  two point function
$\hat{a}(q^2)\gmn$ into the triangle diagram and integrate over second loop
momentum $\ell_2$. The integration has logarithmic divergent part, 
however, it drops out by the projection operator to $a_\mu$ , Eq.(\ref{eq:projection}).
Final expression is very complex and we do not quote here.
From Eq.(\ref{eq:renorma1}), we can see that there are  denominators having
$\ell_2( \leftarrow q)$. 
So the final expression has 5 integration parameters which run in the interval [0,1]. 
As a sample of calculation, we show sum of  boson loop 
contribution. Fish type diagrams,\{$(WW),(W\chi),(\chi W),(\chi \chi),(c^+c^+),(c^-c^-)$\}
and tad pole type diagrams,$ \{(W),(\chi),(c^+),(c^-)\}$
composing $\Pi(q^2)_{\mu\nu}$.

 In unit of $10^{-11}$ we get,
$-6.1048 \times 10^{-3}$ by this method. On the other hand, we get 
$-6.104 \times 10^{-3}$ by our two-loop formalism with CT-terms. The coincidence is 
quite good. Notice that, in this case the two-loop formalism takes huge cpu-time, especially 
for $(W-W)$ diagram.  Its contribution is around $8.8\times 10^6$ in unit of $10^{-11}$
so that we need almost  15 digits number  to cancel UV-divergence. 
So, the effective method is not only important to check the 
reliability of our general formalism but also is useful to get the numerical result.
In the case of self-energy type diagrams, we can construct effective method in several 
cases, however, for vertex type diagrams, to construct  effective method is complex. 
%%%%%%%%%%%%%%%%%%%%%%%%%%%%%%%%%%%%%%%%%%%%
\nocite{*}
\bibliography{reference}

\end{document}